\documentclass[aps,preprintnumbers,floats, prd, twocolumn, longbibliography, nofootinbib]{revtex4-1}
\usepackage{graphicx}
\usepackage{dcolumn}
\usepackage{bm}
\usepackage{amsmath}
\usepackage{amsthm}
\usepackage{multirow}
\usepackage{enumerate}
\usepackage{amsfonts}
\usepackage{ifthen}
\usepackage{psfrag}
\usepackage{slashed}
\usepackage{hyperref}
\usepackage{gensymb}
\usepackage[utf8]{inputenc}
\usepackage{color}
\usepackage{subfigure}
\usepackage{ulem}
\usepackage[utf8]{inputenc}
\usepackage{float}

\newcommand{\be}{\begin{equation}}
\newcommand{\ee}{\end{equation}}
\newcommand{\bea}{\begin{eqnarray}}
\newcommand{\eea}{\end{eqnarray}}

\begin{document}

\title{Dark Matter Subhalo Evaporation by Coulomb-like Interaction with Galactic Gas}

\author{Yugen Lin$^{1,2}$}
\email{linyugen@ihep.ac.cn}
\author{Yu Gao$^{1}$}
\email{gaoyu@ihep.ac.cn}

\affiliation{$^1$ Key Laboratory of Particle Astrophysics, Institute of High Energy Physics, Chinese Academy of Sciences, Beijing, 100049, China}
\affiliation{$^2$ School of Physical Sciences, University of Chinese Academy of Sciences, Beijing, 100049, China}

\begin{abstract}
Coulomb-like interactions typically has a cross section scales with velocity dependence as $\sigma=\sigma_0 v^{-4}$. The momentum transfer rate between a slightly charged dark matter and ionized particles increases significantly at low velocity, and it produces prominent evaporation effects on small-sized dark matter overdensities. We show that when subhalos encounter the hot gases near the Milky Way's disc, their survival can place stringent limits on Coulomb-like scattering strength. For $M<10^5 M_\odot$ subhalos to survive a kilo-parsec distance from the galactic center, with a dark matter mass in the sub-GeV range, the evaporation limit becomes one order of magnitude stronger than the limits from current cosmic microwave background and baryon acoustic oscillation data. We also interpret our bounds into the electron-recoil direct detection cross section, and show that the evaporation effect can lead to a stronger constraint on Coulomb-like interaction for sub-MeV dark matter in comparison with direct detection experiments.
\end{abstract}

\maketitle


\section{Introduction}
\label{sect:intro}
A large number of cosmological and astrophysical observations indicate the existence of dark matter (DM) which makes up about 85\% of the matter in our Universe. In the standard $\rm \Lambda$CDM cosmology, DM is a cold, collisionless (CDM) particle, which fits well with our current observational constraints~\cite{Bertone:2004pz,Planck:2018vyg}. However, there is still some potential evidence that the CDM model could be incomplete~\cite{Boylan-Kolchin:2011qkt,Bullock:2017xww,Weinberg:2013aya}. Although all the evidence we have for dark matter is derived from its gravitational interactions with ordinary matter, there is also the possibility that dark matter can have other, non-gravitational interactions with Standard Model (SM) particles and cause some observable effects throughout the Universe. Nowadays, many dark matter models that go beyond the CDM paradigm have emerged, which further motivates the exploration of interacting DM(IDM)~\cite{Spergel:1999mh,Tulin:2017ara,Boddy:2022knd}.

In IDM scenarios, the relevant physical process of interaction can be described by the momentum transfer cross section which characterize the efficiency of momentum transfer and largely captured as a power law function of the relative particle velocity $v$~\cite{Sigurdson:2004zp,Dvorkin:2013cea,Slatyer:2018aqg,Bi:2021njb,Becker:2020hzj,
Rogers:2021byl,Liu:2019knx,Driskell:2022pax,Ooba:2019erm,Mahdawi:2018euy,Boddy:2018wzy,Li:2022mdj}. In recent years, DM with the Coulomb-like interactions has been the subject of keen interest due to the anomalous observation by the Experiment to Detect the Global EoR Signature (EDGES)~\cite{Bowman:2018yin,Barkana:2018qrx,Berlin:2018sjs}. In such scenario, the transfer cross section between DM and charged particles is proportional to $v^{-4}$, which can cause effectively energy exchange between DM and baryon at later times and cool down baryon to explain the anomalous 21 cm absorption reported by EDGES collaboration. A subclass of these models includes millicharged DM which can couple to the SM photon and only scatter with charged particles. The effective coupling can be produced via kinetic mixing between a dark photon and SM photon through a dark vector portal~\cite{Dvorkin:2019zdi,Baryakhtar:2020rwy}.

Elastic scattering between DM and charged particles can produce an observable effect throughout the Universe. These interactions can affect the cosmic microwave background (CMB) anisotropies and suppress the matter power spectrum on small scales~\cite{Boehm:2000gq,Chen:2002yh,Boehm:2004th}. Previous work has constrained the momentum-transfer
cross section between DM and protons by using measurements of CMB anisotropies from the Planck satellite~\cite{Gluscevic:2017ywp,Boddy:2018kfv,Xu:2018efh,Ali-Haimoud:2015pwa,Ali-Haimoud:2021lka}. Other observational probes of structure including the measurements of Lyman-$\alpha$ forest power spectrum and the abundance of Milky-Way satellite galaxies have also been used to constrain the DM-proton scattering~\cite{Viel:2013fqw,Irsic:2017ixq,Kennedy:2013uta,DES:2020fxi,Nadler:2019zrb,Maamari:2020aqz}. In this work, we start from a novel point and focus on the evaporation effects of subhalos due to the scattering between DM particles and ionized galactic gas. Considering the transfer cross section of Coulomb-like scattering is proportional to $v^{-4}$, so it can produce prominent evaporation effects at low velocity.

In the late Universe, ionized hotspots re-emerge in inner galactic regions. Our Milky Way is one such massive galaxy where the diffuse gas is predominantly ionized. When subhalos travel through such ionized regions, dark matter particles become gradually heated via soft scattering with ionized gas, to a point that they may escape the subhalo's gravitational binding~\cite{Bi:2021njb}. So the survival of subhalos requires stringent limits on the strength of Coulomb-like scattering. One typical scenario to realize such interaction is the kinetically mixed $U(1)'$ model, where the dark matter particles can develop a tiny electric charge under $U(1)_{\rm EM}$ and their Coulumb-like soft-scattering is mediated by the SM photon. In following sections, we will assume the extra $U(1)'$ mass is heavy~\cite{Fabbrichesi:2020wbt,Gninenko:2012eq,Curtin:2014cca,Chang:2016ntp,DOnofrio:2019dcp} so that the dark matter self-interaction can be negligible, and focus on the photon-mediated DM heating by colliding with the hot galactic gas. We will calculate the collision heating rate and place an upper limit on the DM’s transfer cross section by assuming the survival of subhalos in the ionized galactic region.

\section{Heating via Coulomb-like Scattering}
\label{sect:scattering}
We adopt a phenomenological model~\cite{Fayet:1990wx,Feldman:2007wj} to implement the Coulomb-like interactions, in which the dark matter particle $\chi$ is a Dirac fermion and it interacts with the charged particles, such as the galactic protons, through the kinetic mixing between SM photon and a dark photon. The dark matter's effective interaction terms are
\begin{equation}
\mathcal{L} \supset -e A_\mu \bar{p} \gamma^\mu p+ \epsilon g_\chi A_\mu \bar{\chi} \gamma^\mu \chi,
\end{equation}
where $\epsilon$ is the kinetic mixing parameter between the SM photon and the dark photon, and
$g_\chi$ represents the gauge coupling under the dark $U(1)'$.

The scattering process between dark matter and charged particles give rise to the momentum transfer cross section that characterizes the efficiency of momentum transfer,
\begin{equation}
\sigma_{\mathrm{T}} \equiv \int d \Omega \frac{d \sigma}{d \Omega}(1-\cos \theta),
\label{eq:sigma}
\end{equation}
where $\theta$ is the scattering angle. For Coulomb-like interactions, the integration over Eq.~\ref{eq:sigma} will lead to a logarithmic divergence. To regulate the forward scattering singularity of momentum-transfer integral, we use the Debye screening length to be the maximal impact parameter~\cite{McDermott:2010pa}
\begin{equation}
\lambda_D=\sqrt{\frac{T_p}{e^2 x_e n_p}},
\end{equation}
where $n_p$, $T_p$ and $x_e$ denote the proton number density, proton temperature and ionization fraction. For Coulomb-like interactions, the maximal impact parameter is equivalent to a minimal scattering angle
\begin{equation}
\theta_{\min }=\arctan \frac{\epsilon e g_\chi}{4 \pi \mu_{\chi p} v^2 \lambda_{\mathrm{D}}} \approx \frac{\epsilon e g_\chi}{6 \pi T_p \lambda_{\mathrm{D}}}
\end{equation}
where we use $\epsilon\ll1$ and $\mu_{\chi p} v^2/2\approx 3T_p/2$. $v$ represents the relative velocity between the DM and the proton, and $\mu_{\chi p}=\frac{m_\chi m_p}{ m_\chi+m_p}$ is their reduced mass. In terms of the cutoff scattering angle, the momentum-transfer cross section is written by
\begin{equation}
\begin{aligned}
\sigma_{\mathrm{T}} & = \frac{2 \pi \epsilon^2 \alpha \alpha_\chi}{\mu_{\chi p}^2 v^4} \ln\left[\rm csc^2 \left(\frac{\theta_{\min}}{2}\right)\right] \\
& \approx \frac{2 \pi \epsilon^2 \alpha \alpha_\chi}{\mu_{\chi p}^2 v^4} \ln \left(\frac{9 T_p^3}{4 \pi \epsilon^2 \alpha^2 \alpha_\chi x_e n_p}\right),
\end{aligned}
\label{eq:sigmaT}
\end{equation}
where $\alpha_\chi\equiv g_\chi^2/4\pi$. As the relative velocity only appears as $v^{-4}$ in the pre-factor, we can parameterize the transfer cross-section of the collision between dark matter and proton as $\sigma_T=\sigma_0 v^{-4}$ where $\sigma_0$ represents the velocity-stripped momentum transfer cross section.

When DM collides with ionized gas, the important physical quantity is the energy transfer rate, which needs to be averaged over the velocity distributions of dark matter and hot gas. For gases primarily made of ionized hydrogen, the thermally averaged energy transfer rate of per unit time is given by Ref.~\cite{Dvorkin:2013cea,Munoz:2015bca},
\begin{equation}
\begin{aligned}
\frac{\mathrm{d} \Delta E_p}{\mathrm{~d} t}= & \frac{m_\chi \rho_p}{\left(m_\chi+m_p\right)} \int d^3 v_p f_p\left(v_p\right) \int d^3 v_\chi f_\chi\left(v_\chi\right) \\
& \times \sigma_T\left(\left|\vec{v}_\chi-\vec{v}_p\right|\right)\left|\vec{v}_\chi-\vec{v}_p\right|\left[\vec{v}_{\mathrm{cm}} \cdot\left(\vec{v}_p-\vec{v}_\chi\right)\right]
\end{aligned}
\label{eq:thermal_xsec}
\end{equation}
where $\rho_p$ is gas density and $\sigma_T$ is the transfer cross section. $\vec{v}_{p}$ and $\vec{v}_{\chi}$ denote the  velocities of the proton and dark matter in the galactic frame, and $\vec{v}_{\rm cm}$ is the DM-proton center-of-mass velocity. One can assume DM velocity inside a subhalo follows a Maxwellian distribution,
\be
 f_{\chi}(\vec{v}_\chi)= \frac{1}{n}e^{-|\vec{v}_\chi-\vec{v}_0|^2/ \sigma^2_v}
 \label{eq:DM_velocity}
\ee
where $\vec{v}_0$ represents the subhalo's collective velocity that circulate around the center of galaxy. For protons in hot gas, their velocity follows a Boltzmann distribution that may also have a collective motion velocity $\vec{v}_{p0}$,
\be
 f_{p}(\vec{v}_p)= \frac{1}{n}e^{-m_p|\vec{v}_p-\vec{v}_{p0}|^2/{2 k_B T}}.
 \label{eq:proton_velocity}
\ee
Substituting the transfer cross-section $\sigma_T=\sigma_0 v^{n}$ into Eq.~\ref{eq:thermal_xsec} and integrating out the velocity distributions,
\begin{equation}
\frac{{\rm d} \Delta E_{p}}{{\rm d} t}=\frac{m_{p} \rho_{\chi} \sigma_{0}}{\left(m_{\chi}+m_{p}\right)}\left[a I_{1}(n)+b I_{2}(n)\right]
\end{equation}
with $a=m_\chi/(m_\chi+m_p)$, $b=(T_\chi-T_p)/u_{th}^2(m_\chi+m_p)$. The integrals $I_{1}(n)$ and $I_{2}(n)$ take the form
\begin{equation}
\begin{aligned}
I_{1}(n)&=\frac{v^{n+6}}{(2 \pi)^{1 / 2} u_{\text {th }}^{3}} \int_{-\infty}^{\infty} d x e^{-x^{2} r^{2} / 2} \\
& \times \frac{x(x-1)(n+x+4)|x-1|^{n+3}}{(n+3)(n+5)} \nonumber ,
\end{aligned}
\end{equation}

\begin{equation}
\begin{aligned}
I_{2}(n)&=-\frac{v^{n+6}}{(2 \pi)^{1 / 2} u_{\text {th }}^{3}} \int_{-\infty}^{\infty} d x e^{-x^{2} r^{2} / 2} \\
& \times \frac{x(x-1)^{3}[(n+4) x+1]|x-1|^{n+1}}{(n+3)(n+5)}
\end{aligned}
\end{equation}
where $u_{th}$ and $r$ are respectively defined as $u_{th}^2=\frac{T_p}{m_p}+\frac{T_\chi}{m_\chi}$, $r=\frac{v}{u_{th}}$, and $v$ represents the relative velocity between the subhalo and galactic gas. For our case, taking $n=-4$ and the corresponding heating rate due to protons in gas is
\begin{equation}
\frac{{\rm d} \Delta E_{\chi}}{{\rm d} t}=\frac{m_\chi \rho_p \sigma_0}{\left(m_\chi+m_p\right)^2 \sqrt{2 \pi} u_{\mathrm{th}}}\left[2 \frac{T_p-T_\chi}{u_{\mathrm{th}}^2} e^{-\frac{r^2}{2}}+m_p \frac{F(r)}{r}\right]
\label{eq:heating_rate}
\end{equation}
where the function $F(r)$ is
\begin{equation}
F(r) = \operatorname{erf}\left(\frac{r}{\sqrt{2}}\right)-\sqrt{\frac{2}{\pi}} e^{-r^2 / 2} r.
\end{equation}

DM subhalos at a few kpc from the galactic center typically have $v_0\sim 10^{-4}$ to $10^{-3}$. So the relative velocity $|\vec{v}_{\chi}-\vec{v}_{p}|$ between the subhalo and gas is much larger than the velocity dispersion inside the subhalo and gas. We can take the limit where relative velocity $v$ is dominated by $|\vec{v}_0-\vec{v}_{p0}|$ and the energy transfer rate will become insensitive to the gas temperature. In this case, the heating rate in Eq.~\ref{eq:heating_rate} takes a much simpler expression
\begin{equation}
\frac{ {\rm d} \Delta E_{\chi}}{{\rm d} t} = \frac{m_\chi m_p \rho_p \sigma_0}{\left(m_\chi+m_p\right)^2 v},
\label{eq:heating_rate1}
\end{equation}
where $\rho_p$ is the proton's mass density. The additional heating by colliding with electrons can be obtained by replacing $m_p\rightarrow m_e$ and $\rho_p\rightarrow \rho_e=\rho_p\cdot m_e/m_p$.  

In previous studies, the cosmological constraints on the momentum transfer cross section of DM–proton and DM–electron are often investigated separately. In this paper, we consider the combined contribution to dark matter subhalos evaporation by summing up the contribution to Eq.~\ref{eq:heating_rate1} from both protons and electrons in ionized gases. We will first derive the constrain on $\epsilon^2 \alpha_{\chi}$ for different dark matter masses, and then use Eq.~\ref{eq:sigmaT} to obtain the constraints of DM–proton momentum transfer cross section, which is more convenient to compare with existing limits. Also noted that in Eq.~\ref{eq:heating_rate1}, when $m_\chi$ is much heavier than the proton mass, the momentum transfer efficiency drops significantly, so that we will only focus on the DM mass range below GeV scale where the evaporation effects is prominent.

\section{Subhalos Evaporation limits}
\label{sect:evaporation_limits}
Given previous calculations, the time scale for an average DM particle to be heated to its host subhalo's escaped velocity can be estimated as
\be
\tau_{\rm esc.} = \frac{1}{2} m_\chi\left( v_{\rm esc}^2- v_{\rm rms}^2\right) \cdot \left(\frac{{\rm d} \Delta E_{\chi}}{{\rm d}t} \right)^{-1},
\label{eq:tau_esc}
\ee
where $v_{\rm rms}$ is the root-mean-square velocity of DM inside the subhalo. Long-time stability of subhalos would require $\tau_{\rm esc.}$ to be comparable or longer than the age of galaxy. Our Milky Way's age is around $10^{10}$ year, so we use $\tau_{\rm esc.}=10^{10}$ yr to calculate the evaporation limits of subhalos by collision with hot gas. DM particle's root-mean-square velocity $v_{\rm rms}$ and escaped velocity $v_{\rm esc}$ would depend on the subhalo size. For our galaxy, we use an empirical scaling relation for a particle's velocity dispersion $\delta_v$ from the population of sizable subhalos that contain visible stars~\cite{Walker:2009zp},
\be
\delta_v \approx  {3.9}{~\rm km/s}\left(\frac{M}{10^6 M_{\odot}}\right)^{1/3}.
\label{eq:scaling}
\ee
The scaling relation of velocity dispersion with the $1/3$ power-law of the halo mass is based on the predictions of the virial theorem. Evrard et al.~\cite{Evrard:2007py} also verified this relation with high resolution N-body simulations. The root-mean-square velocity $v_{\rm rms}$ and escaped velocity $v_{\rm esc}$ are related to the velocity dispersion $\delta_v$ by an ${\cal O}(1)$ factor, for a Maxwellian distribution in Eq.~\ref{eq:DM_velocity}, $v_{\rm rms} = 1.73 \delta_v$ and $v_{\rm esc}=2.44 \delta_v$.

\begin{figure}[h]
\begin{center}
\subfigure{\includegraphics[scale=0.65]{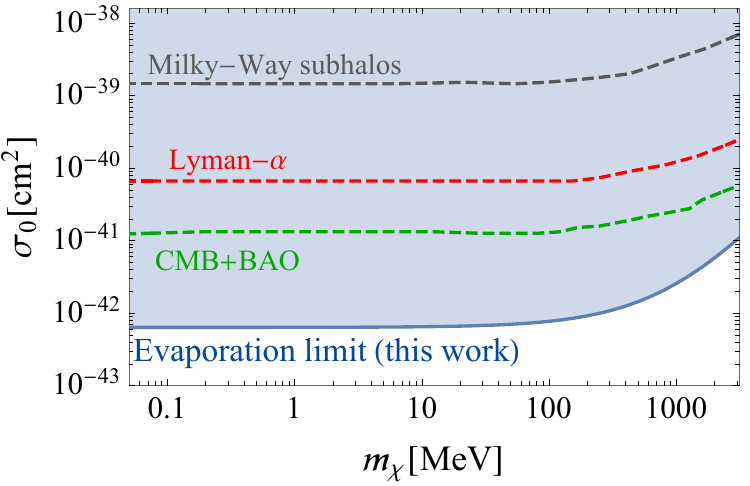}}
\subfigure{\includegraphics[scale=0.65]{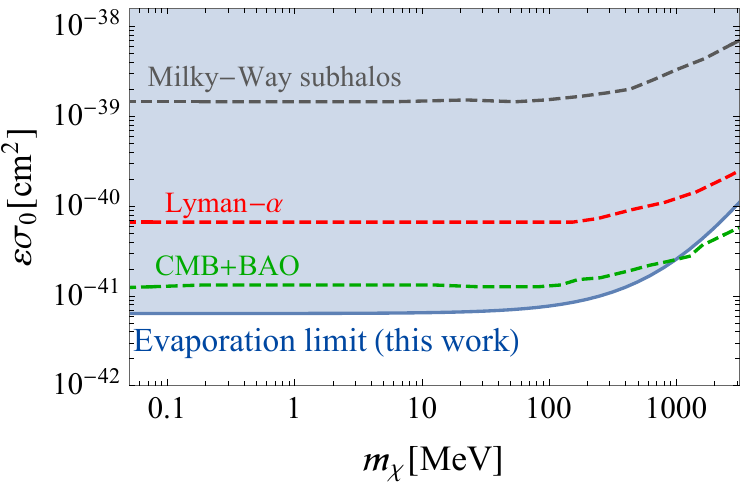}}
\caption{The velocity-stripped momentum transfer cross section $\sigma_0$ limits (solid blue) for $10^5 M_\odot$ subhalos that survive a $10^{10}$ yr evaporation time in an ionized environment with an average particle number density $0.5$ cm$^{-3}$. Subhalo velocity assumes $v=10^{-4}$ at a 1 kpc distance to the galactic center (upper panel) and $v=10^{-3}$ at 10 kpc (lower panel). $\varepsilon$ is time fraction of the subhalo's orbit in the gas-distributed region, and $\varepsilon=1$ for orbits entirely submerged in gaseous regions. The bounds from datasets of CMB+BAO (dashed green), Lyman-$\alpha$ forest (dashed red) and Milky-Way subhalos (dashed gray) are also shown for comparison.}
\label{fig:sigama0}
\end{center}
\end{figure}

Ionized hot clouds are distributed within a few kpc from the Milky Way's disk. The average particle number density is 0.5 cm$^{-3}$ and the temperature is around 8000 K ~\cite{2018ApJ...862...34N}. When subhalos travel through ionized gas near the galactic disk, collisions will heat up DM subhalos towards their binding energy. For convenience of illustration, we choose two orbit radii at 1 kpc and 10 kpc  and the corresponding circulating velocities are $10^{-4}$ and $10^{-3}$ around the center of the galaxy. We use the two benchmark velocities to estimate the evaporation limits. If a subhalo's orbit is not completely submerged in the gaseous region, a fraction of time $\varepsilon$ would apply, and the corresponding evaporation time should scale as $\tau_{\rm esc.}\rightarrow \varepsilon^{-1}\tau_{\rm esc.}$, and the $\sigma_0$ limits scale as $\varepsilon^{-1}$. In the conventional galactic model, hot gas extends to 2 kpc away from the disk, so $\varepsilon$ is between 0.13 and 1 for a 10 kpc circular orbit.

The Milky Way's observable subhalos are typically massive ($M \gtrsim 10^5 M_\odot$) and we use the subhalo mass with $10^5 M_\odot$ to calculate our evaporation limits, while in principle lower-mass subhalo may also form. The evaporation limits for $10^5 M_\odot$ subhalos with $\tau_{\rm esc.}= 10^{10}$ yr are shown as the blue solid line in Fig.~\ref{fig:sigama0}. The upper panel represents the evaporation limits in case of subhalos with velocity $v=10^{-4}$ and a 1 kpc orbit radius, and the lower one represents the situation for subhalos at 10 kpc with a velocity $v=10^{-3}$. In the lower figure, $\varepsilon$ was used at the $\sigma_0$ limit, which represents time fraction of the subhalo's orbit in the gas-distributed region, and $\varepsilon=1$ is corresponding to orbits entirely submerged in gaseous regions. We also shown the constraints from the datasets of CMB+BAO (dashed green), Lyman-$\alpha$ forest (dashed red) and Milky-Way subhalos (dashed gray) for comparison, which are adapted from the recent study Ref.~\cite{Buen-Abad:2021mvc}.

Previous cosmological constraints on the DM-baryon interactions can be briefly summarized as follows: the exchange of heat between DM and baryon would cool the baryons and lead to recombination occurring earlier, which can directly impact the CMB~\cite{Ali-Haimoud:2015pwa,Ali-Haimoud:2021lka}. On the other hand, in the presence of DM-baryon interactions, baryon photon coupling can convey pressure to the dark matter, which damps the clustering of dark matter against structure formation, and leads to a suppression on the matter power spectrum~\cite{Boehm:2000gq,Chen:2002yh,Boehm:2004th}. In this work, we study the evaporation of subhalos via Coulomb-like scattering between dark matter and ionized galactic gas. The subhalos evaporation limits are dependent on the gaseous environment of the inner galaxy and the location of subhalos. In case of $M\le 10^{5} M_\odot$ subhalos would survive at a 1 kpc orbit radius from the center of the Milky Way, the corresponding evaporation limit on $\sigma_0$ is one order of magnitude stronger than previous CMB+BAO constraint for a DM mass range below GeV scale. For a larger 10 kpc orbit radius, the corresponding evaporation limit becomes comparable with the constraint from CMB+BAO.

In Fig.~\ref{fig:sigama0}, it is not surprising that our evaporation limits for near-disk subhalo-gas scattering appears stronger than that from the average subhalo formation in the Milky-Way (gray dashed)~\cite{Buen-Abad:2021mvc}, which derives from the small-scale properties of the matter power spectrum for the extended galactic region and uses the Universe's average baryon number density for energy transfer calculation. As the ionized gas is more concentrated in the inner galactic region, the baryon number density in this region is much larger than the average baryon density of the Universe, which makes collision with ionized gas much more frequent and makes the evaporation limit more prominent for subhalos in this region.

\begin{figure}[tbp]
\begin{center}
\subfigure{\includegraphics[scale=0.65]{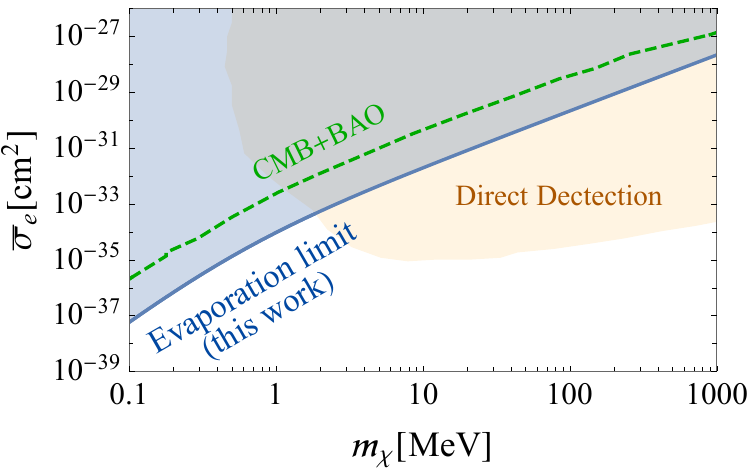}}
\subfigure{\includegraphics[scale=0.65]{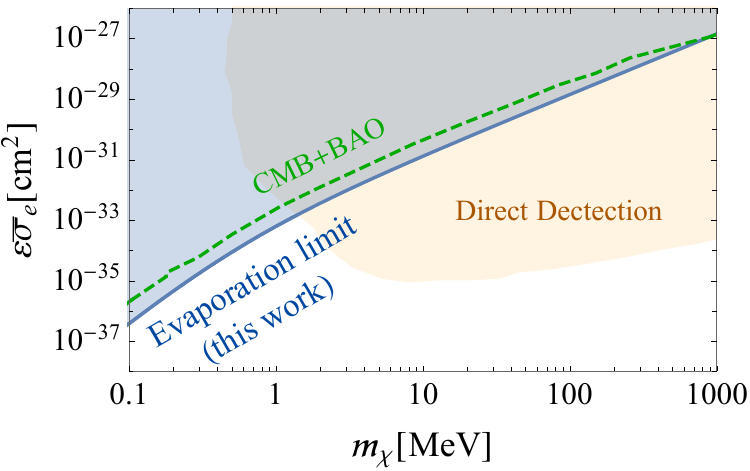}}
\caption{Interpreted evaporation limits (solid blue) into the electron-scattering direct-detection cross section $\bar{\sigma}_e$. Similar to Fig.~\ref{fig:sigama0}, the upper (lower) panel corresponds the evaporation limits of subhalos with velocity $v=10^{-4}$ ($v=10^{-3}$) and a 1 kpc (10 kpc) orbital distance to the galactic center. The combined direct detection exclusion bounds via DM-electron scattering experiments are shown as orange shaped region. The bounds from datasets of CMB+BAO (dashed green) are also shown for comparison.}
\label{fig:sigamabar}
\end{center}
\end{figure}

The direct-detection experiments can also constrain the same $t$-channel scattering process, which can be the complementarity of cosmological probes. Next we will compare our limits to those from direct-detection experiments. For the electronic recoil direct detection experiments, the corresponding cross section is given by ~\cite{Essig:2011nj,Essig:2015cda}
\begin{equation}
\bar{\sigma}_e \equiv \frac{\left.\overline{|\mathcal{M}|^2}\right|_{q=q_{\mathrm{ref}}}}{16 \pi\left(m_e+m_\chi\right)^2}=\frac{\left.\mu_{\chi e}^2 \overline{|\mathcal{M}|^2}\right|_{q=q_{\mathrm{ref}}}}{16 \pi m_\chi^2 m_e^2}
\label{eq:DD}
\end{equation}
where $\overline{|\mathcal{M}|^2}|_{q=q_{\mathrm{ref}}}$ is the spin averaged amplitude-square at a reference transfer-momentum  $q=q_{\mathrm{ref}}$, and $\mu_{\chi e}$ is the DM–electron reduced
mass. A common choice for the reference transfer momentum for DM–electron scattering is $q_{\mathrm{ref}}=\alpha m_e$~\cite{Emken:2019tni}. The $t$-channel spin averaged $|{\cal M}|^2$ for DM-electron scattering is
\begin{equation}
\overline{|\mathcal{M}|^2} = \frac{16 \epsilon^2 e^2 g_\chi^2 m_e^2 m_\chi^2}{t^2}
\label{eq:amplitude}
\end{equation}
where $t$ is the Mandelstam variable. The cross section is
\begin{equation}
\left.\bar{\sigma}_e \equiv \frac{1}{16 \pi\left(m_e+m_\chi\right)^2} \overline{|\mathcal{M}|^2}\right|_{q=q_{\mathrm{ref}}}=\frac{16 \pi \epsilon^2 \alpha \alpha_\chi \mu_{\chi e}^2}{q_{\mathrm{ref}}^4}.
\label{eq:sigmabar}
\end{equation}

One can utilize the relation between DM-proton and DM-electron scattering cross-sections,
\begin{equation}
\bar{\sigma}_e=\frac{\mu_{\chi e}^2}{\mu_{\chi p}^2} \bar{\sigma}_p,
\end{equation}
to relate our velocity-stripped momentum transfer cross section $\sigma_0$ to the direct detection cross section $\bar{\sigma}_e$:
\begin{equation}
\sigma_0=\bar{\sigma}_e \cdot \frac{\alpha^4 m_e^4 \xi}{8\mu_{\chi p}^4},
\end{equation}
where $\xi$ is the Debye logarithm as in Eq.~\ref{eq:sigmaT}. We show the interpreted evaporation limits in terms of $\bar{\sigma}_e$ in Fig.~\ref{fig:sigamabar} to compare with the direct detection results. The solid blue line in upper and lower figure respectively correspond the evaporation limits of subhalos with velocity $v=10^{-4}$ and $v=10^{-3}$ in the galactic frame. The combined direct detection exclusion bounds~\cite{DarkSide:2018ppu,SuperCDMS:2018mne,SENSEI:2020dpa,XENON:2019gfn} are shown as the orange shaded region (right-top) in the ($m_\chi$,$\bar{\sigma}_e$) plane. We also plotted the interpreted CMB+BAO (dashed green) constraints for comparison, which is adapted from Ref~\cite{Buen-Abad:2021mvc}. The bounds from Lyman-$\alpha$ forest and Milky-Way's average subhalos are not shown since they are weaker than those from CMB+BAO. For surviving subhalos located at 1 kpc from the galactic center, and a sub-MeV DM mass, the corresponding evaporation limits exclude a larger region of DM parameter space. This indicates that in case subhalos run into galactic gases, Coulomb-like interaction becomes strongly constrained for DM masses below current direct detection thresholds. We also show the case with a larger orbit radius at 10 kpc, the evaporation limit can also be comparable with the constraint from CMB+BAO.

\section{Summary and Discussion}
\label{sect:conclusion}
In this work, we investigated the evaporation of subhalos through Coulomb-like interaction between dark matter and ionized gas in the inner galactic region. The typical feature for Coulomb-like scattering is that the momentum transfer cross section scales with relative particle velocity as $v^{-4}$,  which renders the collision evaporation very efficient at low velocity. Due to the diffuse gas in our galaxy is mostly ionized and it is concentrated near the galactic disk, for subhalos located in such region, their survival requires stringent limits on the strength of Coulomb-like interactions.

Previous limits on the DM-baryon Coulomb-like interaction mostly derive from the imprints on CMB and structure formation. Soft collision with the fully ionized galactic gases offers a novel evaporation rate limits on such a scattering cross-section. If any $M \le 10^5 M_\odot$ subhalo live around 1 kpc away from the center of the Milky Way, our result can be more stringent than the existing CMB+BAO limit for a DM mass range below the GeV scale. At a larger orbital distance around 10 kpc, such an evaporation limit is still comparable with the CMB and BAO bounds. In comparison with electronic-recoil direct detection constraints, the evaporation limit readily extends to the sub-MeV range that is below current direct detection thresholds and can be stronger than CMB+BAO bounds for subhalos in the inner galactic region. Considering our Milky Way is far from being an active galaxy, the evaporation effects would be further enhanced in galaxies that are more active and have higher amount of fully ionized gas than our Milky Way.

\medskip
{\bf Acknowledgements.}~~\\
The authors acknowledge support from the National Natural Science Foundation of China (12275278).

\bibliography{refs}

\end{document}